\documentclass[twocolumn,prl,amsmath,amssymb,floatfix,superscriptaddress,showpacs]{revtex4-1}
\usepackage{graphicx}
\usepackage{dcolumn}
\usepackage{bm}
\usepackage{xspace}
\usepackage{hyperref}
\usepackage{color}
\usepackage[flushleft]{threeparttable}

\def\yzgo{YbZnGaO$_4$\xspace}
\def\lzgo{LuZnGaO$_4$\xspace}
\def\ymgo{YbMgGaO$_4$\xspace}
\begin{document}

\title{Spin-glass ground state in a triangular-lattice compound \yzgo}
\author{Zhen~Ma}
\author{Jinghui~Wang}
\author{Zhao-Yang~Dong}
\affiliation{National Laboratory of Solid State Microstructures and Department of Physics, Nanjing University, Nanjing 210093, China}
\author{Jun~Zhang}
\affiliation{State Key Laboratory of Surface Physics, Department of Physics, and Laboratory of Advanced Materials, Fudan University, Shanghai 200433, China}
\author{Shichao~Li}
\author{Shu-Han~Zheng}
\affiliation{National Laboratory of Solid State Microstructures and Department of Physics, Nanjing University, Nanjing 210093, China}
\author{Yunjie Yu}
\affiliation{State Key Laboratory of Surface Physics, Department of Physics, and Laboratory of Advanced Materials, Fudan University, Shanghai 200433, China}
\author{Wei~Wang}
\affiliation{National Laboratory of Solid State Microstructures and Department of Physics, Nanjing University, Nanjing 210093, China}
\author{Liqiang~Che}
\affiliation{Center for Correlated Matter and Department of Physics, Zhejiang University, Hangzhou 310058, China}
\author{Kejing~Ran}
\author{Song~Bao}
\author{Zhengwei~Cai}
\affiliation{National Laboratory of Solid State Microstructures and Department of Physics, Nanjing University, Nanjing 210093, China}
\author{P.~\v{C}erm\'{a}k}
\author{A.~Schneidewind}
\affiliation{J\"{u}lich Centre for Neutron Science (JCNS) at Heinz Maier-Leibnitz Zentrum (MLZ), Forschungszentrum J\"{u}lich GmbH, Lichtenbergstr. 1, 85748 Garching, Germany}
\author{S.~Yano}
\affiliation{Neutron Group, National Synchrotron Radiation Research Center, Hsinchu 30077, Taiwan}
\author{J.~S.~Gardner}
\affiliation{Neutron Group, National Synchrotron Radiation Research Center, Hsinchu 30077, Taiwan}
\affiliation{Center for Condensed Matter Sciences, National Taiwan University, Taipei 10617, Taiwan}
\author{Xin Lu}
\affiliation{Center for Correlated Matter and Department of Physics, Zhejiang University, Hangzhou 310058, China}
\affiliation{Collaborative Innovation Center of Advanced Microstructures, Nanjing University, Nanjing 210093, China}
\author{Shun-Li~Yu}
\email{slyu@nju.edu.cn}
\affiliation{National Laboratory of Solid State Microstructures and Department of Physics, Nanjing University, Nanjing 210093, China}
\affiliation{Collaborative Innovation Center of Advanced Microstructures, Nanjing University, Nanjing 210093, China}
\author{Jun-Ming Liu}
\affiliation{National Laboratory of Solid State Microstructures and Department of Physics, Nanjing University, Nanjing 210093, China}
\affiliation{Collaborative Innovation Center of Advanced Microstructures, Nanjing University, Nanjing 210093, China}
\author{Shiyan Li}
\email{shiyan\_li@fudan.edu.cn}
\affiliation{State Key Laboratory of Surface Physics, Department of Physics, and Laboratory of Advanced Materials, Fudan University, Shanghai 200433, China}
\affiliation{Collaborative Innovation Center of Advanced Microstructures, Nanjing University, Nanjing 210093, China}
\author{Jian-Xin Li}
\email{jxli@nju.edu.cn}
\author{Jinsheng Wen}
\email{jwen@nju.edu.cn}
\affiliation{National Laboratory of Solid State Microstructures and Department of Physics, Nanjing University, Nanjing 210093, China}
\affiliation{Collaborative Innovation Center of Advanced Microstructures, Nanjing University, Nanjing 210093, China}


\begin{abstract}
We report on comprehensive results identifying the ground state of a triangular-lattice structured \yzgo to be spin glass, including no long-range magnetic order, prominent broad excitation continua, and absence of magnetic thermal conductivity. More crucially, from the ultralow-temperature a.c. susceptibility measurements, we unambiguously observe frequency-dependent
peaks around 0.1~K, indicating the spin-glass ground state. We suggest this conclusion to hold also for its sister compound \ymgo, which is confirmed by the observation of spin freezing at low temperatures. We consider disorder and frustration to be the main driving force for the spin-glass phase.
\end{abstract}


\maketitle
Quantum spin liquids (QSLs) represent a novel state of matter in which spins are highly entangled, but neither order nor freeze at low temperatures\cite{Anderson1973153,nature464_199}. There is accumulating experimental evidence suggesting that such a state is realized in \ymgo~(refs~\onlinecite{sr5_16419,prl115_167203,np13_117,nature540_559,PhysRevLett.117.097201,PhysRevLett.118.107202,arXiv:1704.06468,arXiv:1708.06655,arXiv:1708.07503}).  The magnetic specific heat $C_{\rm m}$ is proportional to $T^\alpha$ with $\alpha\approx2/3$~(refs~\onlinecite{sr5_16419,PhysRevLett.117.267202,np13_117}). It has a negative Curie-Weiss temperature of $\Theta\sim-4$~K~(refs~\onlinecite{sr5_16419,prl115_167203}) but does not show a long-range magnetic order at low temperatures\cite{nature540_559,np13_117}. Moreover, diffusive continuous magnetic excitations have been observed by inelastic neutron scattering (INS) measurements\cite{nature540_559,np13_117}, which are interpreted as resulting from the fractional spin excitations of a QSL~(refs~\onlinecite{np12_942,nature492_406}). However, there are also reports challenging this idea: i) The thermal conductivity ($\kappa$) study in ref.~\onlinecite{PhysRevLett.117.267202} reveals no contributions to $\kappa$ from magnetic excitations despite the large magnetic specific heat at low temperatures, casting doubts on the existence of itinerant quasiparticles expected for a QSL~(ref.~\onlinecite{Yamashita1246}); ii) Since Mg$^{2+}$ and Ga$^{3+}$ in the nonmagnetic layers are randomly distributed\cite{sr5_16419,prl115_167203,nature540_534}, the disorder effect, which is detrimental to the QSL phase for this compound\cite{PhysRevLett.119.157201}, can be significant\cite{PhysRevLett.118.107202,np13_117}.  

In this Letter, we report comprehensive measurements on a closely related system, \yzgo. We show that the most natural conclusion, that is consistent with the micro- and macro-scopic data presented here is that the system is a spin glass. We suggest this conclusion to be also true for \ymgo, further supported by the observation of spin freezing at low temperatures. We believe  disorder\cite{nature540_534,PhysRevLett.118.107202,PhysRevLett.119.157201,np13_117,arXiv:1708.07503} and frustration\cite{prl115_167203,arXiv:1608.06445,2016arXiv161203447L,arXiv:1705.05699,np13_117,PhysRevB.96.075105,PhysRevB.94.035107,arXiv:1708.07503} to be largely responsible for this phase.

High-quality single crystals of \yzgo were grown by the floating-zone technique, overcoming the problem caused by the volatile nature of ZnO~(refs~\onlinecite{Kimizuka1982166,sr5_16419}). The d.c. and a.c. magnetic susceptibility and specific heat were measured in a Quantum Design physical property measurement system (see Supplementary Materials~\cite{sm} for details). INS experiments on the single crystals were carried out on PANDA located at MLZ at Garching, Germany\cite{panda}. In the measurements, the 11 co-aligned single crystals weighed 1.2~g in total with a sample mosaic of 0.98$^\circ$ were mounted in the $(H,\,K,\,0)$ plane. INS experiments on the 14-g powder sample were carried out on SIKA located at ANSTO at Lucas Heights, Australia. The wave vector $\bm{Q}$ was expressed as ($H,\,K,\,L$) reciprocal lattice unit (r.l.u.) of $(a^{*},\,b^{*},\,c^{*})=(4\pi/\sqrt{3}a,\,4\pi/\sqrt{3}b,\,2\pi/c)$, with $a=b=3.414(2)$~\AA\, and $c=25.140(2)$~\AA.   

\begin{figure}[htb]
\centerline{\includegraphics[width=3.3in]{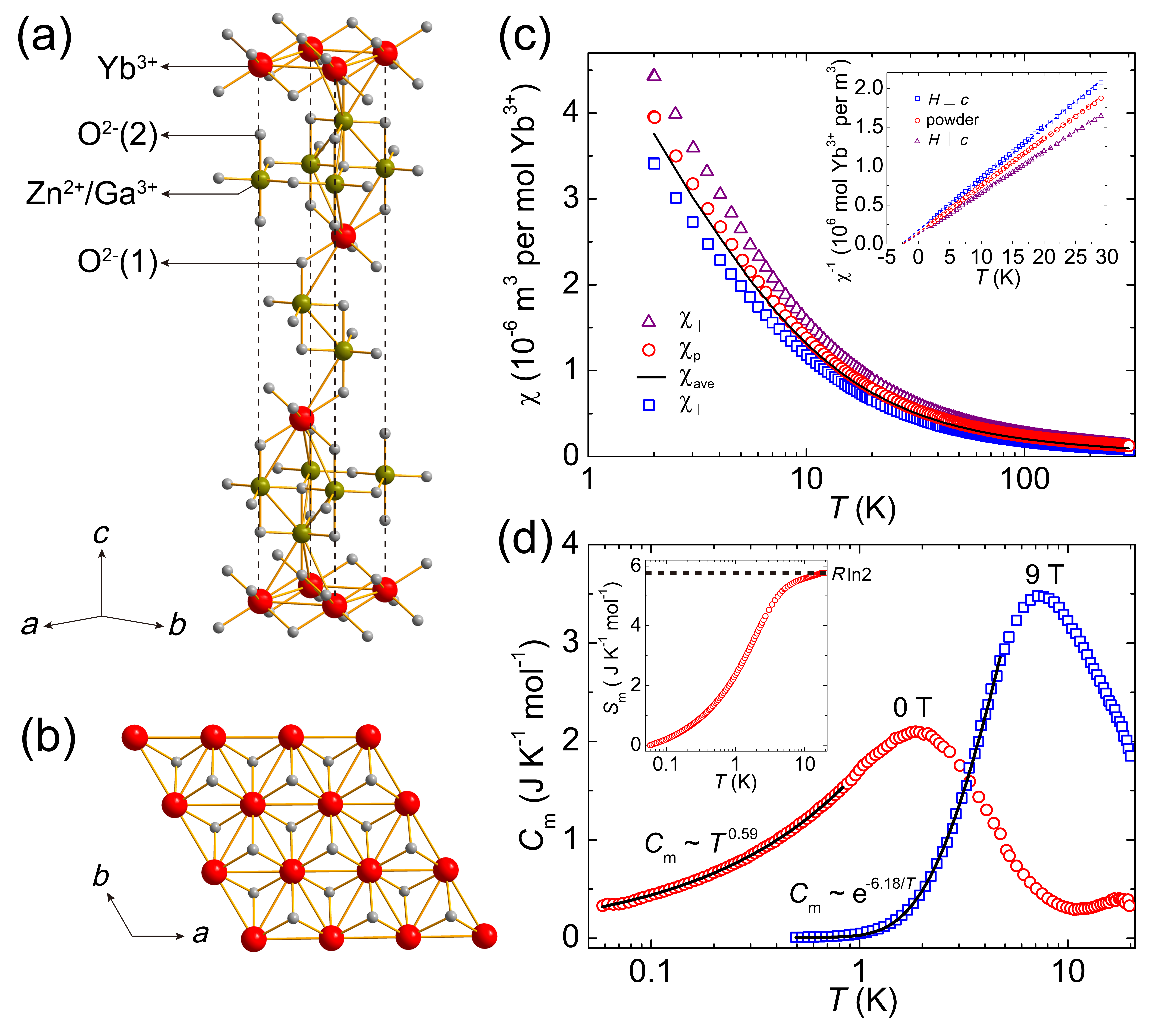}}
\caption{
(a) Schematic crystal structure for \yzgo and \ymgo. (b) Top view of the triangular layer of YbO$_6$ octahedra. (c) D.C. magnetic susceptibility for fields applied parallel ($\chi_{||}$) and perpendicular ($\chi_\perp$) to the $c$ axis for the single crystal, and polycrystalline sample ($\chi_{\rm{p}}$), measured with a 0.1-T field. The data have been corrected by the Van Vleck paramagnetic susceptibility, as discussed in~\cite{sm}. The solid line is the averaged susceptibility $\chi_{\rm{ave}}$. The inset shows the inverse susceptibility and their accompanying Curie-Weiss fits. (d) Magnetic specific heat ($C_{\rm{m}}$) measured at zero and 9-T fields. $C_{\rm{m}}$ is obtained by subtracting the contribution from the lattice using a nonmagnetic reference sample \lzgo. Solid lines are fits to the data described in the main text. The inset shows the zero-field magnetic entropy $S_{\rm m}$, obtained using $S_{\rm m}=\int_0^TC_{\rm m}/TdT$. The dashed line indicates $R$ln2 ($R$, the ideal gas constant).
\label{fig1}}
\end{figure}

\yzgo is isostructural to \ymgo, both of which have the YbFe$_2$O$_4$-type structure (space group $R\bar3m$, No.~166)~(refs~\onlinecite{CAVA1998337,Kimizuka1982166,PhysRevB.61.1811}). Schematics of the crystal structure and two-dimensional triangular lattice of Yb$^{3+}$ are illustrated in Fig.~\ref{fig1}(a) and (b), respectively. The magnetic ground state of Yb$^{3+}$ ions is a spin-1/2 Kramers doublet~(see Fig.~\ref{fig1}d or refs~\onlinecite{sr5_16419,np13_117,PhysRevLett.118.107202}). In \yzgo, the d.c. magnetic susceptibility of the effective spin follows the Curie-Weiss law from 2 to 30~K. In the inset of Fig.~\ref{fig1}(c), we show the inverse susceptibility and the Curie-Weiss fits up to 30~K. From the fits, we find $\Theta$ to be -2.70(2), -2.38(3), and -2.46(2)~K, for the single crystal with magnetic fields perpendicular and parallel to the $c$ axis and for the polycrystalline sample, respectively. The negative sign shows that the magnetic ground state is dominated by antiferromagnetic interactions. The superexchange coupling constant $J$ is estimated to be 1.73(5)~K. These parameters are summarized in Table~\ref{tab:para}, together with other values for this material, in comparison with \ymgo.

\begin{table*}[htb]
  \begin{threeparttable}
\caption{Some parameters for \yzgo and \ymgo.}
\label{tab:para}
\begin{tabular*}{\textwidth}{@{\extracolsep{\fill}}cccccccccccccc}
\hline \hline
\begin{minipage}{2cm}\vspace{1mm} Compound \vspace{1mm} \end{minipage} & $\Theta_\perp$ (K) & $\Theta_{||}$ (K) & $\Theta_{\rm{p}}$ (K)  & $J$ (K) & $g_\perp$ & $g_{||}$ & $g_{\rm{p}}$ & $T^+$ (K) & $\alpha$ & $\Delta$ (K) & $T_{\rm f}$ (K) & $\Delta P$\\
\hline
\begin{minipage}{2cm}\vspace{1mm} \yzgo \vspace{1mm} \end{minipage} & -2.70(2) & -2.38(3) & -2.46(2) & 1.73(5) & 3.17(4) & 3.82(2) & 3.58(3) & 1.86(5) & 0.59(2) & 6.18(3) & 0.093(6) & 0.053(2) \\
 \ymgo & -4.78 & -3.20 & -4.11 & 1.5* & 3.00 & 3.82 & 3.21 & 2.40 & 0.74 &  8.26 & 0.099(6) & 0.068(4) \\
\hline \hline
\end{tabular*}
\begin{tablenotes}
      \small
      \item \quad $\Theta_{\perp}$, $\Theta_{||}$, and $\Theta_{\rm{p}}$ are Curie-Weiss temperatures for the single crystal with magnetic fields perpendicular and parallel to the $c$ axis, and for the polycrystalline sample, respectively. $J$ is the superexchange coupling constant, approximated by\cite{prl115_167203} $J=(4J_{\pm}+J_{zz})/3$, where $J_{\pm}\approx-\Theta_{\perp}/3=0.90(1)$~K, and $J_{zz}\approx-2\Theta_{||}/3=1.59(2)$~K. $g_{\perp}$, $g_{||}$, and $g_{\rm{p}}$ are Land\'{e} $g$ factors, obtained by fitting the magnetization data in Fig.~S2. $T^+$ is the peak temperature of the zero-field magnetic specific heat. $\alpha$ is the fitted index using $C_{\rm{m}}\sim T^\alpha$. $\Delta$ is the magnon gap obtained by fitting the 9-T data with $C_{\rm m}\sim\exp(-\Delta/T)$. Corresponding values for \ymgo are from refs~\onlinecite{nature540_559,sr5_16419,PhysRevLett.117.267202,prl115_167203}.\\
\quad$T_{\rm f}$ is the peak temperature of the real part of the a.c. susceptibility ($\chi^\prime$) at 100~Hz. $\Delta P$ is the peak shift ${\Delta T_{\rm f}\over T_{\rm f}\Delta\lg(f)}$ from 100 to 10000~Hz. These two parameters for both \yzgo and \ymgo are obtained from our own measurements.\\
\quad *Note that in our discussions, we take $J$ to be 2.83~K for \ymgo, larger than the 1.5~K reported in ref.~\onlinecite{prl115_167203}. The larger $J$ is obtained by calculating with the Curie-Weiss temperatures given in ref.~\onlinecite{nature540_559}.
\end{tablenotes}
\end{threeparttable}
\end{table*}

\begin{figure*}[htb]
\centerline{\includegraphics[width=7in]{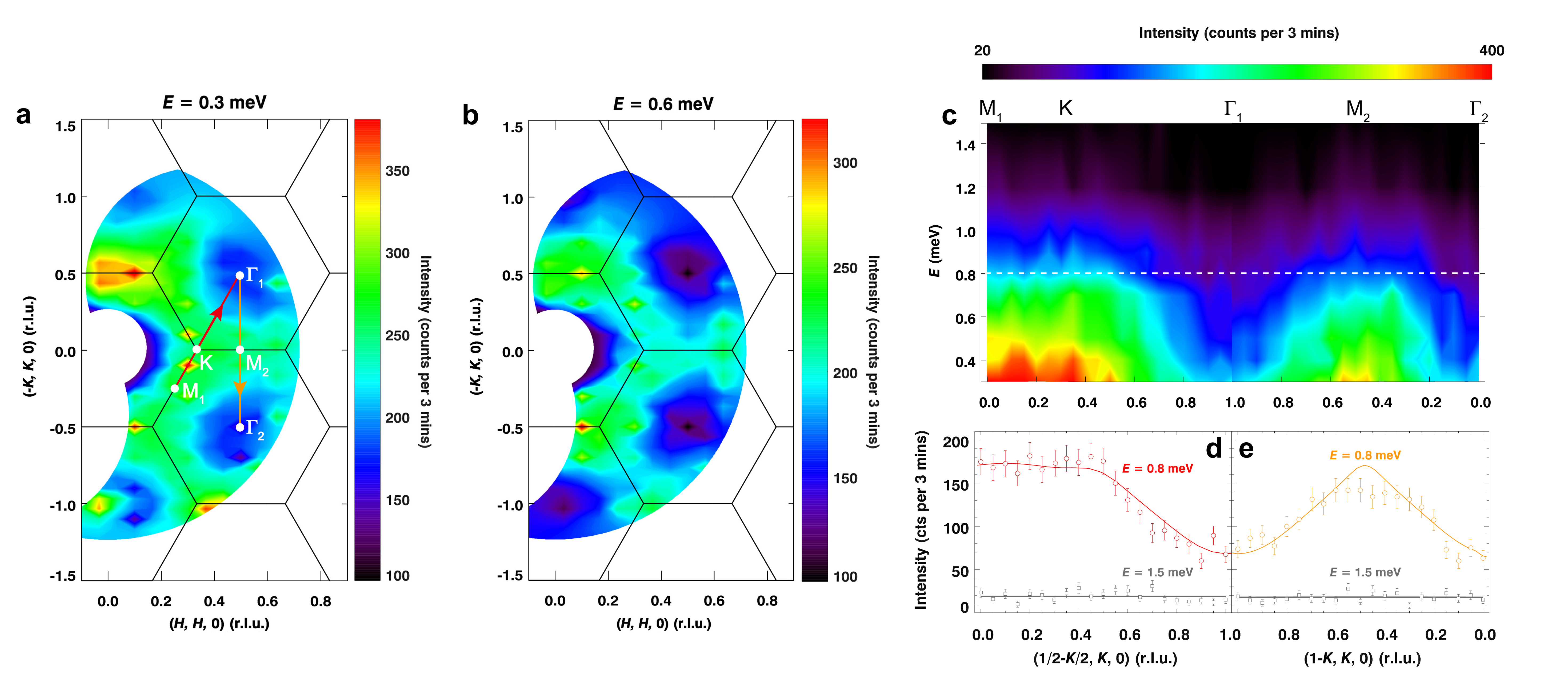}}
\caption{(a) and (b) are contour maps of the INS spectra at $E=0.3$ and 0.6~meV, respectively, measured at $T=0.47$~K. The maps are obtained by plotting together a series of constant-energy scans along the [$H,\,0,\,0$] direction with a step size of 0.1~r.l.u., and an interval of 0.1~r.l.u. along the [$0,\,K,\,0$] direction. Solid lines indicate Brillouin zone boundaries. The additional bright feature around (-0.5,\,0.5,\,0) in (a) does not represent a magnetic Bragg peak~\cite{sm}. (c) Magnetic dispersion along M$_1$-K-$\Gamma_1$ and $\Gamma_1$-M$_2$-$\Gamma_2$ directions as illustrated by the arrows in (a). The dispersion is obtained by plotting together a series of constant-energy scans as shown in (d) and (e), with an energy interval of 0.1~meV. The dashed line indicates constant-energy scans at $E=0.8$~meV. In (d) and (e), lines through data are the calculated results extracted from Fig.~\ref{fig3}. Errors represent one standard deviation throughout the paper.
\label{fig2}}
\end{figure*}

In Fig.~\ref{fig1}(d) we plot the magnetic specific heat ($C_{\rm m}$) down to 0.05~K for \yzgo. From the zero-field data, we do not observe a $\lambda$-type peak expected for a well-defined phase transition. Instead, there is a broad peak at $T^+\approx1.86$(5)~K, below which $C_{\rm m}$ decreases. Below $T^+$, we fit $C_{\rm m}$ to $T^\alpha$ and determine $\alpha$ to be 0.59(2). We have also attempted to fit the low-temperature data using $C_{\rm{m}}\sim\exp(-\Delta/T)$, and obtained a small gap of 0.05(3)~K, consistent with the large magnetic specific heat arising from the gapless magnetic excitations at low temperatures. With increasing fields, $T^+$ gets higher, and the hump becomes narrower. At 9~T, $T^+$ should correspond to a transition from the paramagnetic to ferromagnetic state, as the system is in a fully polarized state at low temperatures [see Fig.~S2(a)]. When we fit the 9-T data with $C_{\rm{m}}\sim\exp(-\Delta/T)$, we obtain $\Delta=6.18(3)$~K, which corresponds to a magnon gap induced by an external magnetic field, as also observed in \ymgo~(refs~\onlinecite{np13_117,PhysRevLett.117.267202}). In the inset of Fig.~\ref{fig1}d, we show that the magnetic entropy $S_{\rm m}$ is precisely $R$ln2 ($R$, the ideal gas constant) at 20~K, expected for a Kramers doublet in the ground state\cite{prl115_167203,np13_117,sr5_16419}.

We now explore the system by carrying out INS experiments, which reveal similar behaviors to \ymgo~(refs~\onlinecite{nature540_559,np13_117}). In Fig.~\ref{fig2}(a) and (b), we present the contour maps of the excitation spectra at energy transfers of $E=0.3$ and 0.6~meV, respectively. The broad diffusive excitations spreading along edges of the two-dimensional (2D) Brilouin zone, and decreasing in intensity with increasing ${\bm Q}$, indicate that the system is dominated by antiferromagnetic correlations but without long-range order, consistent with the macroscopic results in Fig.~\ref{fig1}(c) and (d). Magnetic dispersions along two high-symmetry paths are plotted in Fig.~\ref{fig2}(c), which exhibit a continuum over the whole energy range measured. The excitations are gapless (see also, Fig.~S4), consistent with the specific heat data. Constant-energy scans along M$_1$-K-$\Gamma_1$ at two representative energies are shown in Fig.~\ref{fig2}(d); similar scans along $\Gamma_1$-M$_2$-$\Gamma_2$ are shown in Fig.~\ref{fig2}(e). At $E=0.8$~meV, intensities remain roughly constant from M$_1$ to (1/4,\,1/2,\,0), and then decrease as ${\bm Q}$ approaches $\Gamma_1$. The scan along the $\Gamma_1$-M$_2$-$\Gamma_2$ direction results in a broad peak centering at M$_2$, and the spin-spin correlation length is estimated to be 3~\AA{} from this scan. This length scale is close to those obtained in \ymgo~(refs~\onlinecite{nature540_559,np13_117}) and other QSL candidates\cite{np12_942,nature492_406}.

\begin{figure}[htb]
  \centering
  \includegraphics[width=3.5in]{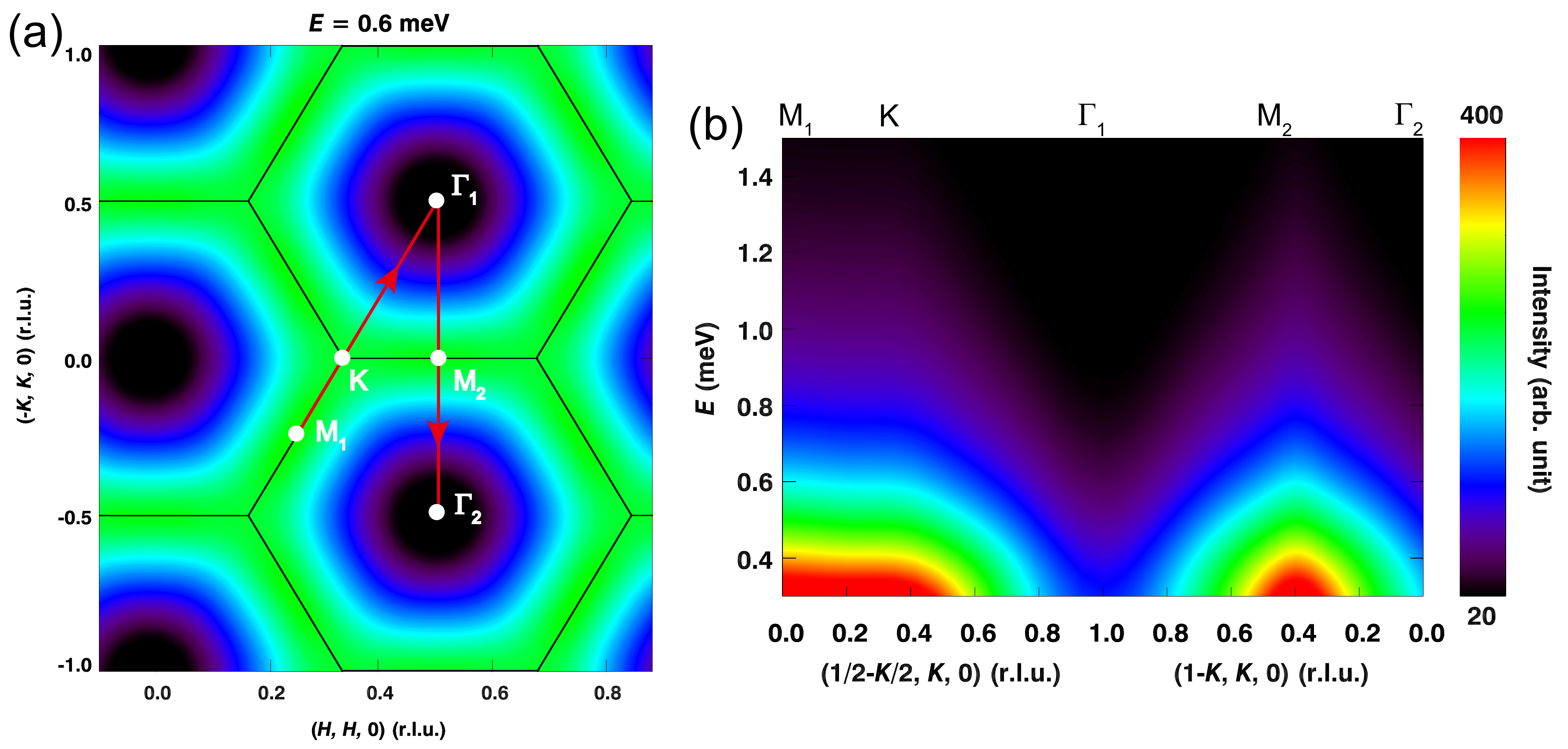}
  \caption{(a) Contour map of the calculated spectra at $E=0.6$~meV. (b) Calculated dispersions along the two high-symmetry paths illustrated in (a).}\label{fig3}
\end{figure}

As is shown in refs~\onlinecite{nature540_559,np13_117}, a QSL phase can give rise to the observed INS spectra. However, we notice that the cations in the nonmagnetic layers are randomly distributed\cite{sr5_16419,prl115_167203,nature540_534,PhysRevLett.118.107202,arXiv:1708.07503}. As a result, there should be a strong variation in the magnetic couplings due to the disordered charge environment\cite{PhysRevLett.119.157201,PhysRevLett.118.107202}. In addition, the small $J$ will further exaggerate the disorder effect.
Can disorder make the magnetic excitations mimic those expected for a QSL~(ref.~\onlinecite{PhysRevLett.119.157201})? In this context, we consider introducing disorder into a stripe-order phase, which is suggested to be the ground state for \ymgo in the absence of disorder\cite{PhysRevLett.119.157201,PhysRevB.95.165110}. We use an anisotropic spin model with nearest-neighbor and next-nearest-neighbor exchange interactions, which has been justified in refs~\onlinecite{PhysRevLett.119.157201,np13_117}, and perform calculations with the linear spin-wave theory\cite{0953-8984-27-16-166002}. Without disorder, gapless spin-wave excitations disperse up from the M point. With increasing disorder, the well-defined spin-wave dispersions become broader both in momentum and energy. An example is presented in Fig.~\ref{fig3}, and the calculated intensities are plotted together with the experimental data in Fig.~\ref{fig2}(d) and (e). The calculated results agree with the experimental data quite well, demonstrating that an antiferromagnet with disorder can also exhibit the continuum-like INS spectra.

\begin{figure}[htb]
\centerline{\includegraphics[width=3.3in]{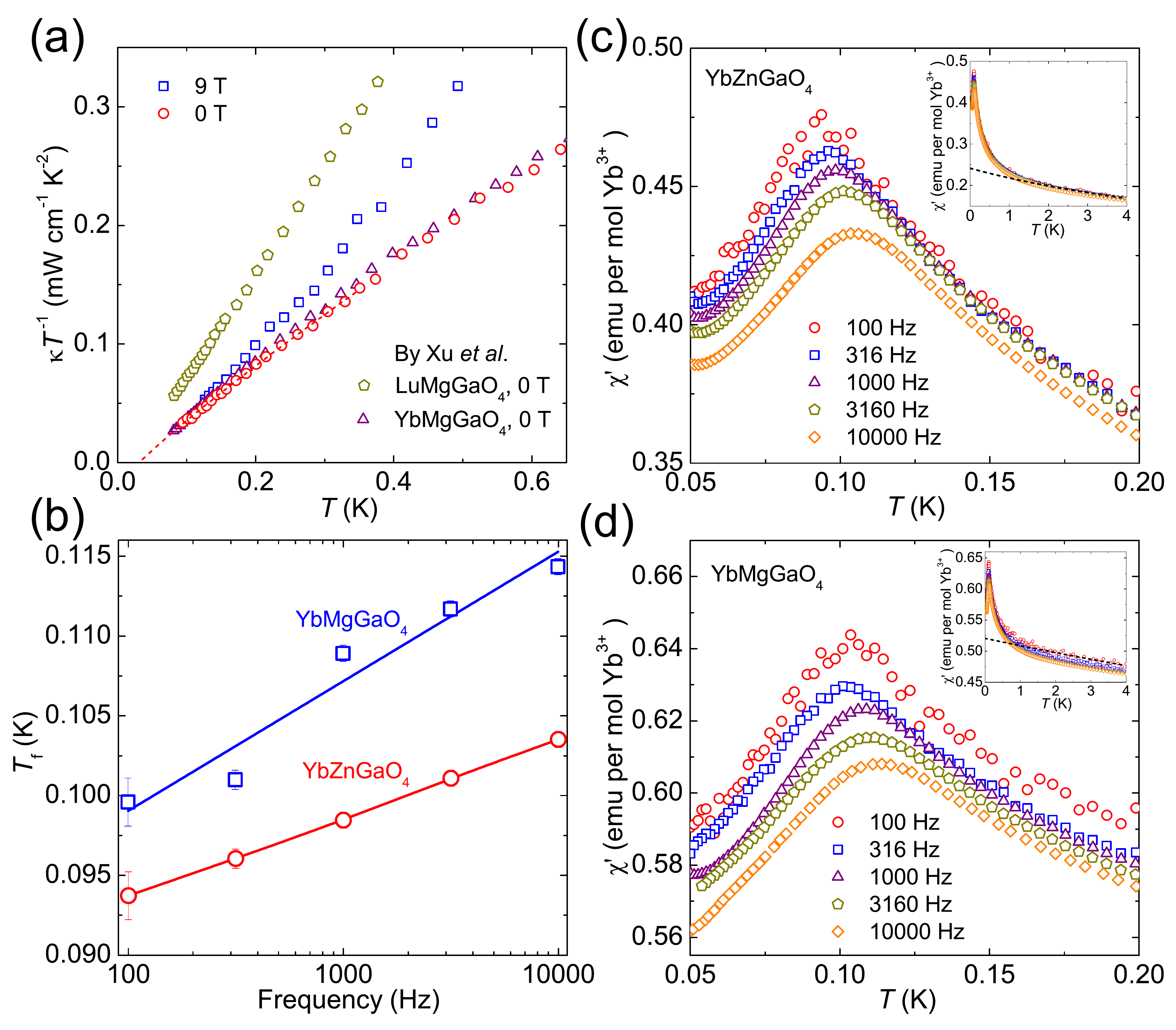}}
\caption{(a) Thermal conductivity results on \yzgo under zero and 9-T magnetic fields applied parallel to the $c$ axis. The dashed line is a fit to the data described in the main text. For comparison, results on \ymgo and the nonmagnetic reference compound LuMgGaO$_4$ are also plotted\cite{PhysRevLett.117.267202}. (b) Frequency dependence of the freezing temperature for both \yzgo and \ymgo, extracted from the temperature dependence of the real part of the a.c. susceptibility ($\chi^{\prime}$) shown in (c) and (d). Lines through data are guides to the eye. In the insets of (c) and (d), $\chi^{\prime}$ in an extended temperature range up to 4~K are plotted. Dashed lines indicate the Curie-Weiss fits for the 100-Hz data.
\label{fig4}}
\end{figure}

We further show thermal conductivity ($\kappa$) results in Fig.~\ref{fig4}(a). At $T=0.1$~K, $\kappa$ is only about half of that of the nonmagnetic sample, LuMgGaO$_4$, in which only phonons contribute to $\kappa$. This reduction is quite likely due to the scattering of phonons off the gapless magnetic excitations\cite{PhysRevLett.117.267202}. This also manifests itself in the magnetic-field measurements: in a field of 9~T that opens a gap of 6.18(3)~K, there are almost no magnetic excitations to scatter phonons, so $\kappa$ increases. We fit the zero-field data with $\kappa/T=\kappa_0/T+nT^{\beta-1}$ up to 0.4~K. Here, the first term $\kappa_0$ and second term $nT^\beta$ represent non-phonon and phonon contributions, respectively. From the fit, we obtain $\kappa_0/T=-0.011(2)$~mW\,K$^{-2}$\,cm$^{-1}$, and $\beta=1.97(2)$. In the nonmagnetic sample LuMgGaO$_4$, it is shown that $\kappa_0/T=-0.007$~mW\,K$^{-2}$\,cm$^{-1}$, and $\beta=2.07$~(ref.~\onlinecite{PhysRevLett.117.267202}). In both materials, $\kappa_0/T$ is virtually zero, similar to the case of \ymgo~(ref.~\onlinecite{PhysRevLett.117.267202}). In contrast, another QSL candidate EtMe$_3$Sb[Pd$(dmit)_2]_2$ has a high $\kappa_0/T=2$~mW\,K$^{-2}$\,cm$^{-1}$, considered to be a signature of  highly mobile quasiparticles in the QSL state\cite{Yamashita1246}. We therefore believe that a gapless QSL is not an applicable description for \yzgo, because its magnetic excitations will contribute to $\kappa$~(refs~\onlinecite{PhysRevB.76.235124,PhysRevB.72.045105}). On the other hand, the thermal conductivity results can be understood within a disordered-magnet picture, in which the mean-free path of the magnons is reduced with disorder, and they are not expected to conduct heat.

Taking all aforementioned observations into account, we believe that \yzgo is a spin glass, with frozen, short-range correlations below the freezing temperature $T_{\rm f}$~(refs~\onlinecite{RevModPhys.58.801,Mydosh1993,Mydosh1986}). Such a phase can be identified from the a.c. susceptibility. Thus, we perform the measurements with temperatures spanning about 3 decades, ranging from 0.05 to 4~K. The results are shown in Fig.~\ref{fig4}(c). At a measuring frequency of 100~Hz, the real part of the susceptibility $\chi^\prime$ shows a broad peak at $T_{\rm f}\approx0.093(6)$~K. The peak height decreases, and the peak temperature increases, with increasing driving frequency $f$. The frequency dependence of $T_{\rm f}$ is shown in Fig.~\ref{fig4}(b). As a quantitative measure, $\Delta P={\Delta T_{\rm f}\over T_{\rm f}\Delta\lg(f)}$ is 0.053(2) with $f$ changing from 100 to 10000~Hz. This value is close to those observed in other insulating spin glasses\cite{PhysRevB.27.3100,RevModPhys.58.801,PhysRevB.92.134412,Mydosh1993,Mydosh1986}. The strong frequency dependence evidences a broad distribution of the spin relaxation times around $T_{\rm f}$, typical for a spin glass\cite{RevModPhys.58.801,Mydosh1993,Mydosh1986}. 

We also measure the a.c. susceptibility for \ymgo. As shown in Fig.~\ref{fig4}(d), the behaviors are similar to those of \yzgo, albeit with a slightly higher $T_{\rm f}$ of 0.099(6)~K at 100~Hz. For \ymgo, the peak shift from 100 to 10000~Hz, $\Delta P$ is 0.068(4)~[Fig.~\ref{fig4}(b)], larger than that for \yzgo. In the insets of Fig.~\ref{fig4}(c) and (d), we plot $\chi^\prime$ in the whole temperature range measured. At high temperatures, it follows the Curie-Weiss law. Below $\sim$2~K, it rises more rapidly with cooling. Remarkably, this temperature coincides with $T^+$, below which the magnetic specific heat decreases.

The spin-glass phase identified from the a.c. susceptibility is a natural ground state for \yzgo: i) Disorder and frustration, the two ingredients for a spin glass\cite{RevModPhys.58.801,Mydosh1993,Mydosh1986,0295-5075-93-6-67001}, are present and strong in such materials\cite{nature540_534,PhysRevLett.118.107202,PhysRevLett.119.157201,np13_117,prl115_167203,arXiv:1608.06445,2016arXiv161203447L,arXiv:1705.05699,PhysRevB.96.075105,PhysRevB.94.035107,arXiv:1708.07503,PhysRevB.96.094414,0953-8984-25-35-356002};  ii) A spin glass maintains short-range spin-spin correlations\cite{RevModPhys.58.801,Mydosh1993,Mydosh1986}, consistent with the absence of a long-range magnetic order; iii) The observed INS spectra can be nicely reproduced by bringing disorder into an ordered state; iv) Macroscopically, a spin glass is disordered, and thus the magnons do not conduct heat due to the short mean-free path. This explains the lack of contribution to the thermal conductivity from the gapless magnetic excitations; v) Finally, we estimate the fraction of frozen moment to be 13(3)\% from our INS results shown in Fig.~S4, close to the 16(3)\% in \ymgo~(ref.~\onlinecite{np13_117}), but smaller than the 33\% expected from theory\cite{PhysRevB.93.014433}. We consider it to be a consequence of the strong frustration in this compound. 

We also note that some findings in \yzgo suggest deviations from a generic spin glass. For instance, $T^+$ is about 20 times of $T_{\rm f}$, much larger than that expected for a typical spin glass\cite{RevModPhys.58.801,Mydosh1993,Mydosh1986}. We believe that this indicates the existence of strong frustration. In addition to the geometrical frustration inherent to the triangular structure\cite{Anderson1973153,nature464_199}, the spin-space anisotropy induced by the spin-orbit coupling of the Yb$^{3+}$ ions, recognized in our anisotropic spin model and in refs~\onlinecite{prl115_167203,arXiv:1608.06445,2016arXiv161203447L,arXiv:1705.05699,np13_117,PhysRevB.96.075105,PhysRevB.94.035107,arXiv:1708.07503}, should further reduce $T_{\rm f}$. Moreover, the strong disorder\cite{nature540_534,PhysRevLett.118.107202,PhysRevLett.119.157201,np13_117,arXiv:1708.07503} is expected to result in a reduced $T_{\rm f}$. In some spin glasses, $C_{\rm m}\sim T^\alpha$ with $\alpha=1$ at low temperatures\cite{PhysRevLett.35.1792,PhysRevLett.85.840}, but disorder may reduce this exponent\cite{arXiv:1710.06860}.

In summary, we have successfully grown high-quality single crystals for \yzgo, and our comprehensive measurements on these crystals provide concrete evidence that it is a spin glass. We show this conclusion is  also applicable to \ymgo. We suggest that the spin-glass phase in both compounds is driven by disorder and frustration. Our work reveals the very similar characteristics between QSL and spin-glass phases, including the broad ``continuum" of magnetic excitations and is a cautionary tale about labelling such materials without a full study of the spin system.

We acknowledge the Applications Group at Quantum Design for measuring the a.c. susceptibility. We thank Fengqi~Song and Haijun~Bu for the help in measuring the high-field magnetization. We are grateful for the stimulating discussions with Jia-Wei~Mei, Shao-Chun~Li, Weiqiang~Yu, Lei~Shu, D.~Adroja, Guangyong~Xu, and J.~M.~Tranquada. The work was supported by the National Natural Science Foundation of China with Grants No.~11374143, 11674157, 11774152, 11374138, 11674158, 11374257, and U1630248, and by the National Key Projects for Research \& Development of the Ministry of Science and Technology of China with Grants No.~2016YFA0300401, 2016YFA0300101, and 2016YFA0300503.

Z. M., J. H. W., Z.-Y. D., and J. Z. contributed equally to this work.


%

\end{document}